\documentclass[aps,pre,floatfix]{revtex4}
\usepackage{epsf}
\usepackage{subfigure}
\usepackage{amsmath}
\usepackage{amssymb}
\usepackage{graphicx}
\usepackage{epstopdf}
\begin{document}

\newcommand{\be}{\begin{equation}}
\newcommand{\ee}{\end{equation}}
\newcommand{\bea}{\begin{eqnarray}}
\newcommand{\eea}{\end{eqnarray}}
\newcommand{\mean}[1]{\left \langle #1 \right \rangle}

\title{Signatures of classical bifurcations in the \\ quantum scattering resonances of dissociating molecules}

\author{Pierre Gaspard}
\affiliation{Center for Nonlinear Phenomena and Complex Systems,\\
Universit\'e Libre de Bruxelles, Code Postal 231, Campus Plaine, 
B-1050 Brussels, Belgium}

\begin{abstract}
A study is reported of the quantum scattering resonances of dissociating molecules
using a semiclassical approach based on periodic-orbit theory.
The dynamics takes place on a potential energy surface with an energy barrier
separating two channels of dissociation.
Above the barrier, the unstable symmetric-stretch periodic orbit may
undergo a supercritical pitchfork bifurcation, leading to a classically chaotic regime.
Signatures of the bifurcation appear in the spectrum of resonances, which have
a shorter lifetime than classically expected.
A method is proposed to evaluate semiclassically the energy and lifetime
of the quantum resonances in this intermediate regime.
\end{abstract}

\maketitle

\section{Introduction}
\label{intro}

It is a great privilege and pleasure to contribute to this special volume
 in honor of Gregory Ezra who has so many
theoretical achievements in the classical and quantum dynamics of molecules.
Methods of dynamical systems theory and semiclassical quantization have proved
to be very powerful to understand the rovibrational dynamics of stable and dissociating molecules.
If the former are characterized by discrete energy spectra in their ground state, the latter have continuous spectra with possible resonances.  Quantum resonances appear at specific energies and their width is inversely proportional to their lifetime.  They manifest themselves in collisions, as well as in dissociations, and are thus called scattering resonances.  They play a fundamental role in our understanding of unimolecular reactions and transition state theory.  

Since the seventies, much effort has been devoted to the semiclassical quantization of classically chaotic dynamics, in particular, using periodic-orbit theory \cite{B81,G90,V88,GR89a,GR89b,GR89c,CE89,ERTW91,GR93,GAB95,GB97}.  These methods have also been applied to scattering processes such as unimolecular reactions \cite{BG94,BG95,BG97}.  Remarkably, the transition state of dissociating molecules is spanned by periodic orbits that are unstable and form a chaotic saddle over large energy ranges.  Yet, bifurcations happen at specific energies where the periodic orbits may become stable and form so-called Kolmogorov-Arnold-Moser (KAM) elliptic islands \cite{AA68}.  Several types of bifurcations have been identified including the supercritical and subcritical pitchfork bifurcations \cite{BG94,BG95,BG97}.  At such bifurcations, the spectrum of quantum scattering resonances undergoes changes, which are difficult to investigate in terms of standard periodic-orbit theory because it fails for stable periodic orbits.

The purpose of the present paper is to show that the semiclassical theory can be extended to obtain the quantum scattering resonances close to pitchfork bifurcations.  The vehicle of our study is the collinear dynamics of HgI$_2$ on the potential energy surface considered by Zewail and coworkers \cite{DBGZ89,GRZ90}.  In previous work \cite{BG94,BG95,BG97}, the classical and quantum dynamics of this system have been analyzed in terms of periodic-orbit theory, showing that the classical dynamics has a single unstable periodic orbit below a critical energy and becomes fully chaotic above.  The quantum scattering resonances have been calculated using standard periodic-orbit theory in the periodic and chaotic regimes.  Here, the semiclassical theory is developed at the pitchfork bifurcation in between both regimes and a local approximation is obtained for the quantum scattering resonances, completing in this way previous work \cite{BG94,BG95,BG97}. Semiclassical theory has the advantage of providing compact analytical expressions describing series of scattering resonances.

This paper is organized as follows.  Section~\ref{sec:2} is devoted to the semiclassical theory in order to determine
the dynamics and the quantum scattering resonances of unimolecular reactions.  In Section~\ref{sec:3}, the semiclassical theory is developed to obtain the quantum scattering resonances close to a supercritical pitchfork bifurcation using the transfer operator method \cite{GR89a,GR89b,GR89c,B92a,B92b}.  In Section~\ref{sec:4}, the theory is applied to the dissociation of HgI$_2$, which manifests such a bifurcation in the classical dynamics of its transition state.  Conclusions are drawn in Section~\ref{sec:5}.

\section{Semiclassical periodic-orbit theory of molecular dissociation}
\label{sec:2}

\subsection{Classical dynamics}

The dissociation of triatomic molecules -- or the reaction between an atom and a diatomic molecule --
takes place on a potential energy surface typically presenting an energy barrier between two valleys, 
in which the molecular fragments separate.  The passage is open above a critical energy
where a saddle point exists.

The classical dynamics of the reaction can be studied in the phase space of atomic positions
and linear momenta.  Since the total energy and the total angular momentum are conserved, the
phase space can be decomposed into shells where the trajectories remain.  Reacting systems
are of scattering type in the sense that the centers of mass of the fragments move in free flight
at a large distance from each other.  The motion is thus unbounded and most of the trajectories
are coming from and running away towards infinity.  Nevertheless, there might exist invariant sets
of bounded trajectories that remain trapped at finite distance.  Such invariant sets may be composed
of a single or infinitely many trajectories.  In the former case, the trajectory is unstable and periodic, 
or reduced to the saddle point.  In the latter case, the invariant set contains
many stable or unstable periodic orbits.  The periodic orbits are stable of elliptic type 
if the invariant set forms KAM islands \cite{AA68}.  
All the periodic orbits are unstable of hyperbolic type if the invariant set is
a fully chaotic saddle.  Remarkably, such chaotic invariant sets may exist over large energy ranges
in the dynamics of dissociating molecules such as HgI$_2$ or CO$_2$ \cite{GB97,BG94,BG95,BG97}.  
These chaotic saddles appear after that the unique unstable periodic orbit existing at energies just above the saddle point has undergone bifurcations.

The classical dynamics can be analyzed in the phase space by taking a Poincar\'e surface of section \cite{W88,W92,N95} in a plane transverse to the symmetric-stretch periodic orbit.  For the collinear dynamics of a triatomic molecule, the phase space has dimension six.  Eliminating the motion of the center of mass and using energy conservation, we remain with a phase space of dimension three, so that the Poincar\'e surface of section is bidimensional and easy to depict.

\begin{figure}
\includegraphics[width=0.7\textwidth]{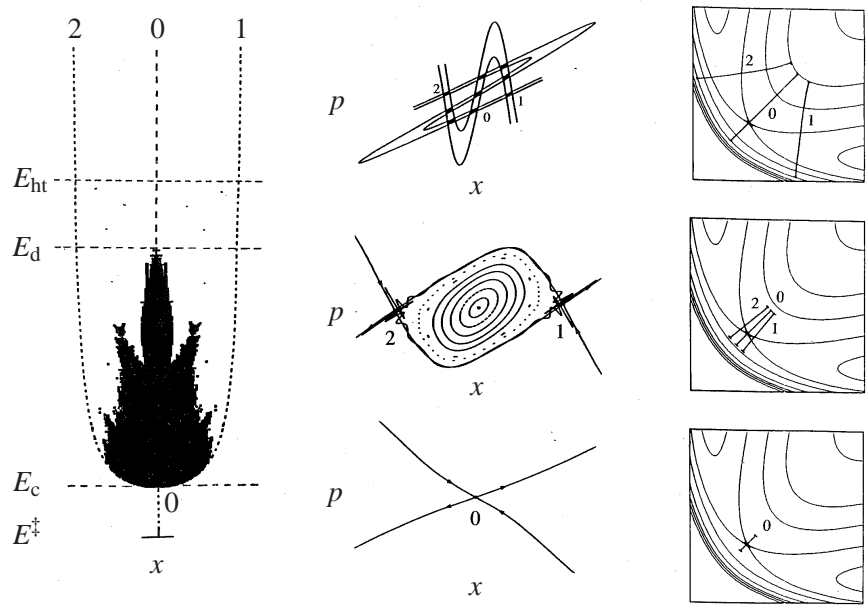}
\caption{Supercritical pitchfork bifurcation scenario for symmetric ABA molecules: On the left, bifurcation diagram in the plane of the energy $E$ versus the position $x$.  The critical energies are, 
respectively, the energy $E^{\ddagger}$ of the saddle point, the energy $E_{\rm c}$ of the pitchfork bifurcation, the energy $E_{\rm d}$ where the symmetric-stretch periodic orbit becomes unstable again, and the energy $E_{\rm ht}$ of the last heteroclinic tangencies, after which the invariant set becomes fully chaotic.  In the center, typical phase portraits in some Poincar\'e surface of section $(x,p)$ in the different regimes.  On the right, the shortest periodic orbits in position space $(r_{\rm AB},r_{\rm BA})$. This scenario is observed to occur in the classical dynamics of HgI$_2$ \cite{GB97,BG94,BG95}.}
\label{fig1}
\end{figure}

Figure~\ref{fig1} shows the bifurcation scenario in symmetric molecules ABA, such as HgI$_2$, where a supercritical pitchfork bifurcation happens.  Below the bifurcation, the unstable periodic orbit corresponds to the symmetric-stretch motion of the triatomic molecule.  This periodic orbit becomes stable of elliptic type at the bifurcation where two new unstable periodic orbits appear corresponding to asymmetric-stretch movements.  Thereafter, the symmetric-stretch periodic orbit is surrounded by a KAM elliptic island, which extends between the two new unstable periodic orbits.  As energy increases, the island itself undergoes successive bifurcations, which destabilize the orbits.  After the last heteroclinic tangencies, the invariant set becomes a fully chaotic saddle composed of unstable orbits of hyperbolic type.  In this chaotic regime, all the orbits are in one-to-one correspondence with bi-infinite sequences
\be
\pmb{\omega} = \cdots\omega_{-2}\omega_{-1}\cdot\omega_0\omega_1\omega_2\cdots
\label{sym-dyn}
\ee
made of three possible symbols $\omega_k\in\{0,1,2\}$.  The symbol $\omega_k=0$ is associated with
a passage close to the symmetric-stretch periodic orbit and the symbols $\omega_k=1$ or $2$
with a passage near by either one or the other of the two asymmetric-stretch periodic orbits.
It is fascinating to observe that this symbolic dynamics captures all the trapped trajectories over a very large
range at high energies.

\begin{figure}
\includegraphics[width=0.7\textwidth]{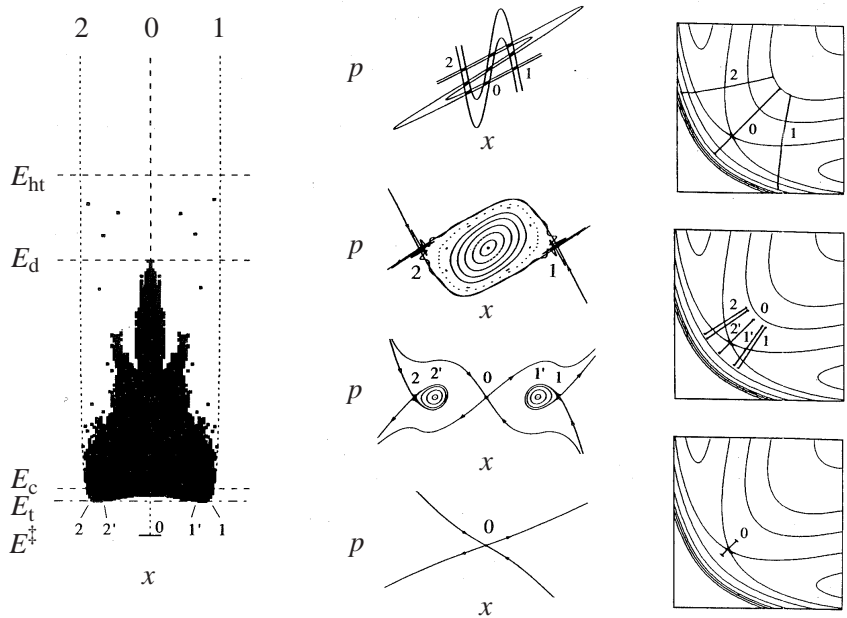}
\caption{Subcritical pitchfork bifurcation scenario for symmetric ABA molecules, represented by the same diagrams as in Fig.~\ref{fig1}. Here, $E_{\rm t}$ denotes the critical energy of two saddle-center tangent bifurcations preceding the subcritical pitchfork bifurcation at $E_{\rm c}$.  This scenario is observed to occur in the classical dynamics of CO$_2$ \cite{GB97,BG97}.}
\label{fig2}
\end{figure}

Figure~\ref{fig2} depicts the scenario in other symmetric molecules ABA, such as CO$_2$, where the pitchfork bifurcation is subcritical and preceded by two saddle-center tangent bifurcations giving birth to a pair of small KAM islands merging together at the subcritical pitchfork bifurcation.  Again, the main KAM island is
bordered by the two unstable periodic orbits of asymmetric-stretch type and the scenario at high energy is
similar to the previous one with a fully chaotic saddle based on three symbols.

\begin{figure}
\includegraphics[width=0.7\textwidth]{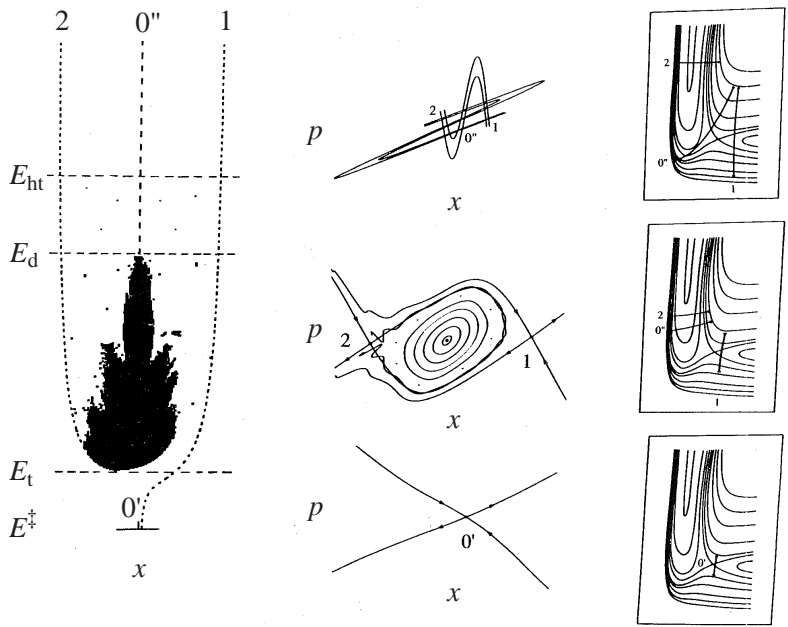}
\caption{Saddle-center tangent bifurcation scenario for nonsymmetric ABC molecules, represented by the same diagrams as in Fig.~\ref{fig1}. Here, $E_{\rm t}$ denotes the critical energy of the saddle-center tangent bifurcation.}
\label{fig3}
\end{figure}

In the case of a nonsymmetric molecule ABC, a different scenario is observed.  Since the symmetry is broken,
the pitchfork bifurcation is replaced by a saddle-center tangent bifurcation giving birth to the main KAM island,
while the unstable periodic orbit continues to exist from lower to higher energies, as shown in Fig.~\ref{fig3}.

An important remark is that the main KAM elliptic island has a small phase-space extension in light-heavy-light molecules, but a larger extension in heavy-light-heavy ones such as the HCl$_2$ or HI$_2$ radicals.

Already the classical dynamics allows us to estimate the lifetime of dissociating molecules by considering the
escape of a statistical ensemble of trajectories from the region near the saddle point.  The number $N_t$ of
trajectories that are still found in this region at a given time $t$ is decaying at a rate, which defines a unimolecular reaction rate as
\be
k_t \equiv -\frac{1}{N_t} \, \frac{dN_t}{dt} \, .
\ee
If the classical dynamics features a KAM elliptic island, the decay of $N_t$ is known to be non exponential \cite{GR89JPC}.  In particular, $N_t$ may converge to a non-vanishing value in the long-time limit if the elliptic island has a positive volume in a three-dimensional energy shell.  Most remarkably, the decay is exponential if the invariant set of trapped trajectories contains a single hyperbolic periodic orbit or is a fully chaotic saddle. In such generic circumstances, a unimolecular reaction rate characterizing the long-time decay can be rigorously defined already in the classical dynamics: $\gamma_{\rm cl}=\lim_{t\to\infty} k_t$.  This rate is equal to the so-called escape rate of dynamical systems theory \cite{ER85}.  It can be expressed as the difference between the positive Lyapunov exponent $\lambda$ measuring the sensitivity to initial conditions and the Kolmogorov-Sinai (KS) entropy per unit time $h_{\rm KS}$ characterizing the randomness of the chaotic dynamics in the invariant set \cite{ER85,KG85}
\be
\gamma_{\rm cl} = \lambda-h_{\rm KS} \, .
\label{esc}
\ee
These quantities vary with energy.  If the invariant set contains a single periodic orbit, the KS entropy vanishes, $h_{\rm KS}=0$, and the unimolecular reaction rate is thus given by the positive Lyapunov exponent: $\gamma_{\rm cl}=\lambda$.  More generally, the classical decay of the number $N_t$ can be decomposed in terms of exponential decays associated with the so-called Pollicott-Ruelle resonances of the classical dynamics \cite{GAB95,GB97}.  

We notice that the invariant set also controls the dynamics of binary reactions.  In the periodic regime,
the stable and unstable manifolds of the single unstable periodic orbit constitute phase-space barriers to the reaction \cite{PP78}.  Similar phase-space barriers have been recently discovered, which are associated with higher index saddles \cite{CEW11}.  If the saddle is chaotic, its stable and unstable manifolds form a fractal set, hence the complexity of the classical reaction dynamics.

\subsection{Quantum dynamics}

Of course, all these classical results have to be reconsidered at the light of quantum mechanics.
The sharp classical trajectory structures are replaced by smooth wavefunction features.
Every classical structure with a phase-space area smaller than Planck's constant $\hbar$
has no quantum correspondence.  Only the classical structures extending in phase space
over scales larger than Planck's constant are susceptible to emerge out of quantum-mechanical waves.
Accordingly, the KAM elliptic islands generated in the aforementioned bifurcation scenarios
are smoothed out if they have an area smaller than Planck's constant.
This is the case for light-heavy-light dissociating molecules such as HgI$_2$ and CO$_2$ where we should
thus expect that the quantization can be carried out semiclassically with a few classical orbits approximating the dynamics over scales larger than Planck's constant $\hbar$.

Quantum mechanically, the configuration of a triatomic molecule is described by a wavefunction $\psi({\bf r}_1, {\bf r}_2, {\bf r}_3, t)$ ruled by Schr\"odinger's equation
\be
i \, \hbar \, \partial_t  \psi = \hat H \, \psi \, ,
\ee
where $\hat H$ is the Hamiltonian operator.  For a dissociating molecule, the energy spectrum is continuous, $\hat H\phi_E=E\phi_E$, extending upwards from the lowest zero-point energy $E=0$ in the valleys of dissociation, so that a wave packet can be decomposed as
\be
\psi_t = \int_0^{\infty} dE \, c(E) \, {\rm e}^{-iEt/\hbar} \, \phi_E
\label{decomp-cont}
\ee
with some function $c(E)$ determined by the initial wave packet $\psi_0$.
At energies above the saddle point, the spectrum may feature resonances corresponding to complex energies, $E_n={\cal E}_n-i\Gamma_n/2$, with an imaginary part giving their width $\Gamma_n$ and lifetime $\tau_n=\hbar/\Gamma_n$.  These resonances can be obtained as generalized eigenstates of the Hamiltonian operator
\be
\hat H \, \phi_n = E_n \, \phi_n = \left({\cal E}_n-i\,\frac{\Gamma_n}{2}\right) \phi_n \, ,
\ee
allowing us to decompose the forward time evolution of a wave packet into a sum of exponentially decaying functions:
\be
\psi_t = \sum_n c_n \ {\rm e}^{-i\left({\cal E}_n-i\Gamma_n/2\right)t/\hbar} \, \phi_n + O(t^{-\nu})
\label{decomp-res}
\ee
up to a possible power-law decay \cite{GAB95}.  The expansion~(\ref{decomp-res}) is obtained by deforming the integration path of Eq.~(\ref{decomp-cont}) from real to complex energies in order to pick up the exponentially decaying contributions of the resonances and the algebraic tail $O(t^{-\nu})$ due to the branch point at $E=0$ \cite{T00,AB81}.  The magnitude of this latter is proportional to the probability amplitude of the initial wave packet at $E=0$, which describes the quasi free flight of the molecular fragments in the dissociation valleys \cite{R65}.

The quantum scattering resonances can be obtained semiclassically using Gutzwiller's trace formula if all the periodic orbits in the invariant set of trapped trajectories are unstable of hyperbolic type \cite{G90}.  As previously shown \cite{GR89a,GR89b,GR89c,CE89,ERTW91,GR93,GAB95,GB97,BG94,BG95,BG97}, the scattering resonances are thus given as the zeroes at complex energy of the Gutzwiller-Voros zeta function
\be
Z(E) = \prod_{m=0}^{\infty} \prod_p \left[ 1 - \frac{{\rm e}^{\frac{i}{\hbar}S_p(E)-i\frac{\pi}{2}\mu_p}}{\vert\Lambda_p(E)\vert^{1/2}\Lambda_p(E)^m}\right]=0 \, ,
\label{Zeta}
\ee
which is a product over all the prime periodic orbits $\{p\}$, i.e., the periodic orbits that are not repeating themselves in phase space \cite{G90,V88}.  These orbits are characterized by their reduced action
\be
S_p(E) = \oint_p {\bf p}\cdot d{\bf q} \, ,
\ee
their Maslov index $\mu_p$, and their instability factor $\Lambda_p(E)$ with respect to the linearized classical dynamics.  The period of each orbit is given by
\be
T_p(E) = \frac{d}{dE} S_p(E) \, ,
\ee
and the Lyapunov exponent characterizing the sensitivity to initial conditions in the vicinity of the periodic orbit by
\be
\lambda_p(E) = \frac{1}{T_p(E)} \, \ln \vert\Lambda_p(E)\vert \, ,
\ee
which is positive for unstable periodic orbits of hyperbolic type.  

It turns out that it is enough to consider the leading factor with $m=0$ in the infinite product over the integer $m$ in order to obtain good semiclassical approximations for the main quantum scattering resonances.  In particular, for an invariant set composed of a single unstable periodic orbit, the main scattering resonances are given by the zeroes of the unique factor with $m=0$ in the zeta function (\ref{Zeta}).  Using the relation $\exp(2\pi i n)=1$ for any integer $n$, the scattering resonances are given by the complex zeroes of
\be
S(E) = 2\pi\hbar \left( n + \frac{\mu}{4}\right) -i \, \frac{\hbar}{2} \, \ln \vert\Lambda(E)\vert  \, ,
\label{quantiz-per}
\ee
with $n=0,1,2,...$ and $\mu=2$ for the symmetric-stretch periodic orbit. Since the imaginary part of the energy is usually smaller than its real part, the action can be expanded around the energy of the resonances, ${\rm Re}\, E_n={\cal E}_n$ satisfying $S({\cal E}_n)=2\pi\hbar ( n + 1/2)$, so that the widths of the resonance are determined by the Lyapunov exponent of the periodic orbit: $\Gamma_n=\hbar\lambda({\cal E}_n)$.  Consequently, in this periodic regime where the KS entropy vanishes $h_{\rm KS}=0$, the lifetimes $\tau_n=\hbar/\Gamma_n=1/\lambda({\cal E}_n)$ of the quantum scattering resonances coincide with the classical lifetime given by the inverse of the escape rate (\ref{esc}): $\tau_{\rm cl}=1/\gamma_{\rm cl}=1/\lambda$.

At energies very close to the critical energy $E^{\ddagger}$ of the saddle point, we notice that the reduced action can be approximated by a linear function of the energy as $S(E)\simeq T^{\ddagger} (E-E^{\ddagger})$.  Replacing in Eq.~(\ref{quantiz-per}), the scattering resonances are estimated to be
\be
E_n \simeq E^{\ddagger} + \hbar\omega^{\ddagger} \left( n + \frac{1}{2}\right) -i \, \frac{\hbar}{2} \, \lambda^{\ddagger}  \, ,
\label{quantiz-harm}
\ee
with $n=0,1,2,...$ and the angular frequency $\omega^{\ddagger}=2\pi/T^{\ddagger}$ in the symmetric-stretch direction of the saddle point.  Accordingly, we recover the expression expected from the harmonic approximation at the saddle point.

If the invariant set is fully chaotic and in correspondence with the triadic symbolic dynamics (\ref{sym-dyn}), semiclassical approximation for the resonances can be obtained as the zeroes of
\be
1 -\sum_{p\in\{0,1,2\}} \frac{{\rm e}^{\frac{i}{\hbar}S_p(E)-i\frac{\pi}{2}\mu_p}}{\vert\Lambda_p(E)\vert^{1/2}} \simeq 0 \, ,
\label{Zeta-3}
\ee
the higher order terms being negligible in the zeta function (\ref{Zeta}) if the periodic orbits are unstable enough.  In this regard, the resonances are determined by interferences between the quantum amplitudes associated with the three shortest periodic orbits $p\in\{0,1,2\}$, which describe the symmetric stretch $p=0$ with $\mu_0=3$ and the asymmetric stretches $p=1,2$ with $\mu_1=\mu_2=2$.  In the chaotic regime where $h_{\rm KS}\neq 0$, the quantum lifetimes may be longer than the classically expected lifetime $\tau_{\rm cl}=1/\gamma_{\rm cl}$ given in terms of the escape rate~(\ref{esc}).  This lengthening is due to quantum interferences described by Eq.~(\ref{Zeta-3}) between the multiple periodic orbits composing the chaotic saddle.  Such interferences do not exist in the periodic regime where the quantum and classical lifetimes thus coincide. In both the classical and chaotic regimes, the spectrum of quantum scattering resonances is gapped as ${\rm Im}\, E_n=-\Gamma_n/2 \leq -\hbar\gamma_{\rm q}/2$ by the quantum escape rate defined as
\be
\gamma_{\rm q} = \lambda^{(1/2)} - 2 \, h_{\rm KS}^{(1/2)} \, 
\label{esc-q}
\ee
where the superscript $(1/2)$ means that the Lyapunov exponent and the KS entropy are defined with respect to a probability distribution where the orbits are weighted by the absolute value of their quantum amplitude, $\vert\Lambda_p(E)\vert^{-1/2}$ \cite{GR89a,GR89b,GR89c,GAB95,GB97}.  In comparison to the classical escape rate~(\ref{esc}), Eq.~(\ref{esc-q}) has an additional factor of $2$ multiplying the KS entropy because classical dynamics evolves probabilities given by squaring the quantum amplitudes.  In the periodic regime where $h_{\rm KS}=h_{\rm KS}^{(1/2)}=0$, the quantum escape rate reduces to the classical one: $\gamma_{\rm q} = \lambda^{(1/2)} = \lambda=\gamma_{\rm cl}$.

These methods have been applied to obtain the scattering resonances of HgI$_2$ and CO$_2$ \cite{GAB95,GB97,BG94,BG95,BG97}.

\section{Semiclassical quantization near a supercritical pitchfork bifurcation}
\label{sec:3}

At the critical energy $E=E_{\rm c}$ of a supercritical pitchfork bifurcation, the unique unstable periodic orbit
of symmetric-stretch motion becomes stable of elliptic type, so that its Lyapunov exponent vanishes, $\lambda(E_{\rm c})=0$, and its instability factor becomes equal to unity $\vert\Lambda(E_{\rm c})\vert=1$. In this case, the Gutzwiller-Voros zeta function~(\ref{Zeta}) no longer provides a reliable approximation for the quantization and other methods are required.  Local and uniform approximations have been developed for bounded systems with discrete energy spectra \cite{OH87,AE94,SS97}.  Here, we extend these methods to open systems with continuous spectra and resonances.

In order to describe the pitchfork bifurcation of the symmetric-stretch periodic orbit, we consider the quantum analogue of a classical Poincar\'e first-return map in a surface of section transverse to the orbit \cite{GR89b,GAB95,B92a,B92b}.  The quantum transfer operator from and to a section in configuration space can be taken as
\be
\psi_{k+1}(x) = \hat Q \, \psi_k(x) = \int \frac{dx'}{\sqrt{2\pi i\hbar}} \, {\rm e}^{\frac{i}{\hbar}F(x,x')-i\frac{\pi}{2}\mu}  \, \psi_k(x') \, ,
\label{tr-op}
\ee
with the generating function
\be
F(x,x') = \frac{1}{2} \, (x-x')^2 - v(x') + F_0 \, ,
\label{gen-fn}
\ee
and the Maslov index $\mu=2$ because the paths undergo two turning points before their return in the surface of section.  The corresponding classical map can be written as
\be
\left\{
\begin{array}{l}
p_{k+1} = p_k -\frac{dv}{dx}(x_k) \, ,\\
x_{k+1} = x_k + p_{k+1} \, .
\end{array}
\right.
\label{cl-map}
\ee

In order to describe a supercritical pitchfork bifurcation, the effective potential is taken as
\be
v(x) = \frac{\chi}{2}\, x^2 - \frac{\alpha}{4}\, x^4 
\label{eff-pot}
\ee
where $\alpha>0$ is a positive constant coefficient while $\chi(E)$ is a parameter varying with the energy $E$ and changing sign at the bifurcation energy $E=E_{\rm c}$ \cite{GAB95,GB97,BG94,BG95,BG97}.  The classical map (\ref{cl-map}) has thus the following form:
\be
\left\{
\begin{array}{l}
p_{k+1} = p_k -\chi \, x_k + \alpha\, x_k^3 \, ,\\
x_{k+1} = p_k + (1-\chi)\, x_k + \alpha\, x_k^3 \, .
\end{array}
\right.
\label{cl-map-bif}
\ee
We notice that $\alpha$ has the units of the inverse of an action, i.e., the units of~$\hbar^{-1}$.

If $\chi<0$, this map admits a unique unstable fixed point, $x=p=0$, which is characterized by the instability factor
\be
\Lambda_0 = 1 -\frac{\chi}{2} + \sqrt{\frac{\chi^2}{4}-\chi} > 1 \, ,
\ee
and the reduced action $S_0=F_0$.

If $\chi>0$, this map has three fixed points: $x=p=0$ which is now stable of elliptic type and two new satellite fixed points at $p=0$ and $x=\pm\sqrt{\chi/\alpha}$ which are unstable with the instability factors
\be
\Lambda_{1,2} = 1 +\chi + \sqrt{\chi^2+2\chi} > 1 \, ,
\ee
and the reduced actions $S_{1,2}=F_0-\chi^2/(4\alpha)$.

The behavior is similar to the continuous-time classical dynamics of symmetric molecules such as HgI$_2$ and we can thus identify the reduced action and instability factors of the map with those of the continuous-time dynamics.  This allows us to obtain the energy dependence of the parameter $F_0$ in the generating function~(\ref{gen-fn}) from the reduced action of the symmetric-stretch periodic orbit, $F_0(E)=S_0(E)$, and the energy dependence of the parameter~$\chi$ in the effective potential~(\ref{eff-pot}) from the instability factor of this orbit also:
\be
\chi(E) = 2 - \Lambda_0(E) - \Lambda_0(E)^{-1} \, .
\ee 
This parameter vanishes as $\chi(E)\propto E-E_{\rm c}$ at the bifurcation, is negative below, and positive above.  The constant coefficient $\alpha$ is obtained from the difference between the reduced actions of the symmetric- and asymmetric-stretch periodic orbits in the limit where the energy $E$ reaches the critical energy $E_{\rm c}$: 
\be
\alpha = \lim_{E\to E_{\rm c}} \frac{\chi(E)^2}{4\left[ S_0(E)-S_{1,2}(E)\right]} \, .
\label{alpha}
\ee

Now that the suitable energy dependence has been given to the quantum transfer operator~(\ref{tr-op}), the scattering resonances are obtained as the zeroes of the characteristic determinant of this transfer operator:
\be
\det\left[ \hat I - \hat Q(E)\right] = 0 \, ,
\label{det}
\ee
at complex energies $E=E_n$ \cite{GR89b,GAB95}. By using the relation 
\be
\ln\det(\hat I-\hat Q)={\rm tr}\ln(\hat I-\hat Q)\, ,
\ee 
the determinant~(\ref{det}) can be expanded in a series involving the trace of the powers of the transfer operator as
\bea
\det\left( \hat I - \hat Q\right) &=& \exp\left( - \sum_{k=1}^{\infty} \frac{1}{k} \, {\rm tr}\, \hat Q^k\right) \nonumber\\
&=& 1 - {\rm tr}\, \hat Q - \frac{1}{2} \, \left[ {\rm tr}\, \hat Q^2 -\left({\rm tr}\, \hat Q\right)^2\right] + \cdots 
\eea
Let us suppose that the terms get smaller and smaller as their power in $\hat Q$ increases.  The first approximation to consider would thus be
\be
1 - {\rm tr}\,\hat Q(E) \simeq 0 \, .
\label{det-approx}
\ee

Now, the trace of the transfer operator (\ref{tr-op}) can be calculated as
\bea
{\rm tr}\,\hat Q &=&  \int \frac{dx}{\sqrt{2\pi i\hbar}} \, {\rm e}^{\frac{i}{\hbar}F(x,x)-i\frac{\pi}{2}\mu}  \nonumber\\
&=& {\rm e}^{\frac{i}{\hbar}F_0-i\frac{\pi}{2}\mu} \int_{-\infty}^{+\infty} \frac{dx}{\sqrt{2\pi i\hbar}} \, {\rm e}^{-i\,v(x)/\hbar}  \nonumber\\
&=& {\rm e}^{\frac{i}{\hbar}F_0-i\frac{\pi}{2}\mu} \int_{-\infty}^{+\infty} \frac{dx}{\sqrt{2\pi i\hbar}} \, \exp\left(-\frac{i\chi}{2\hbar}\, x^2 +\frac{i\alpha}{4\hbar} \, x^4 \right)  \, ,
\eea
which is known to be given by
\be
{\rm tr}\,\hat Q = {\rm e}^{\frac{i}{\hbar}F_0-i\frac{\pi}{2}\mu} \sqrt{\frac{\pi\chi}{8i\hbar\alpha}}\, {\rm e}^{-\frac{i\chi^2}{8\hbar\alpha}} \left[ {\rm e}^{i\frac{\pi}{8}} J_{-\frac{1}{4}}\left(\frac{\chi^2}{8\hbar\alpha}\right) + {\rm e}^{-i\frac{\pi}{8}} J_{\frac{1}{4}}\left(\frac{\chi^2}{8\hbar\alpha}\right)\right]  \, ,
\ee
in terms of the Bessel functions of fractional order $J_{\pm\frac{1}{4}}(z)$ \cite{AE94,SS97}.

Close to the pitchfork bifurcation, the parameter $\chi(E)$ is vanishing and the Bessel functions can be approximated by their Taylor series
\be
J_\nu(z) \simeq \frac{1}{\Gamma(\nu+1)} \, \left(\frac{z}{2}\right)^{\nu}
\ee
where $\Gamma(\nu+1)$ is the Gamma function of argument $\nu+1$ \cite{AS72}.  Within this approximation, we get
\be
{\rm tr}\,\hat Q = {\rm e}^{\frac{i}{\hbar}F_0-i\frac{\pi}{2}\mu-i\frac{\pi}{8}} \frac{\Gamma(\frac{1}{4})}{2\,\pi^{1/2} \hbar^{1/4}\alpha^{1/4}} + O(\chi)
\ee
where we used the relation $\Gamma(\frac{1}{4})\Gamma(\frac{3}{4})=\pi\sqrt{2}$.  Using the identification of the parameter $F_0$ with the reduced action of the bifurcating symmetric-stretch periodic orbit, $F_0=S(E)$, and $\mu=2$, we obtain the result that the scattering resonances close to the pitchfork bifurcation should be given as the zeroes of
\be
S(E) = 2\pi\hbar \left( n + \frac{1}{2}+ \frac{1}{16}\right) -i \, \hbar \, \ln \frac{2\,\pi^{1/2} \hbar^{1/4}\alpha^{1/4}}{\Gamma(\frac{1}{4})}   \, ,
\label{quantiz-bif-per}
\ee
with $n=0,1,2,...$ and under the condition that $\hbar\alpha > \left[\Gamma(\frac{1}{4})/(2\sqrt{\pi})\right]^4=1.09422...$

\section{The case of HgI$_2$}
\label{sec:4}

The previous results are applied to the dissociation process
\be
h\nu + {\rm HgI}_2(X^1\Sigma_g^+) \to \left[{\rm IHgI}\right]^{\ddagger} \to {\rm HgI}(X^2\Sigma^+) + {\rm I}(^2P_{3/2}) \, ,
\ee
reported by Zewail and coworkers \cite{DBGZ89,GRZ90}. The collinear dynamics is modeled according to the Hamiltonian:
\be
\hat H = \frac{1}{2\mu_{\rm HgI}} \left( \hat p_1^2 + \hat p_2^2\right) - \frac{1}{m_{\rm Hg}} \, \hat p_1\, \hat p_2 + V(r_1,r_2) \, ,
\label{Hamilt-HgI2}
\ee
with the damped Morse potential for two degrees of freedom proposed in Ref.~\cite{GRZ90} and the reduced mass $\mu_{\rm HgI}=(m_{\rm Hg}^{-1}+m_{\rm I}^{-1})^{-1}$.  The origin of the energy scale is taken at the critical energy $E^{\ddagger}=0$ of the saddle point.  Time and energy units are chosen respectively as femtosecond and cm$^{-1}$.  In these units, Planck's constant has the value $\hbar=5308.84$~fs~cm$^{-1}$.

The dynamics of the Hamiltonian system (\ref{Hamilt-HgI2}) has been analyzed with classical, semiclassical, and wavepacket methods in Refs.~\cite{GB97,BG94,BG95}.  This analysis has shown the existence of a supercritical pitchfork bifurcation at the critical energy $E_{\rm c}=523$~cm$^{-1}$ where the single unstable symmetric-stretch periodic orbit becomes stable of elliptic type over the small energy range $E_{\rm c}<E<548$~cm$^{-1}$.  In this range, this orbit is surrounded by a small KAM island, which is bordered by two unstable asymmetric-stretch periodic orbits born at the pitchfork bifurcation.  The KAM elliptic island undergoes further bifurcations until the last heteroclinic tangencies at $E=575$~cm$^{-1}$, above which the invariant set is fully chaotic and described by the triadic symbolic dynamics~(\ref{sym-dyn}).

\begin{figure}
\includegraphics[width=0.5\textwidth]{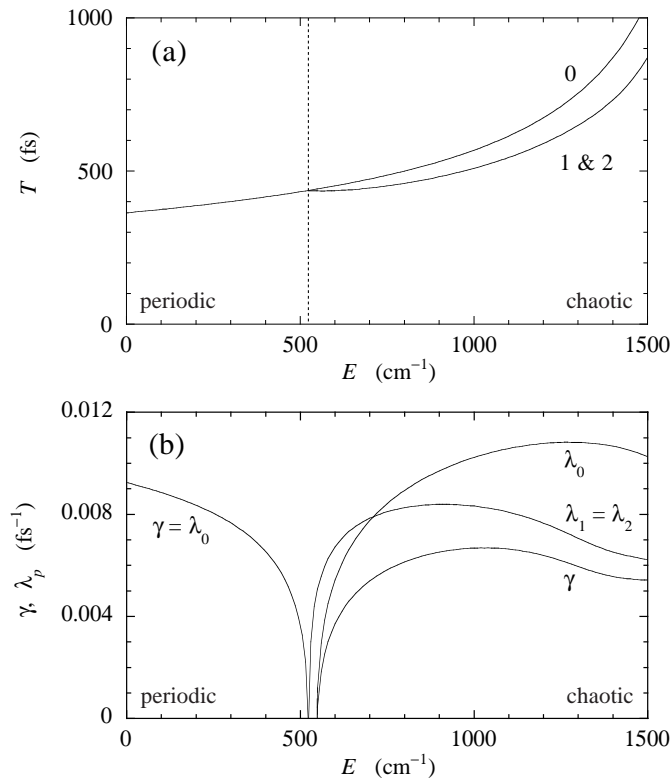}
\caption{Characteristic quantities of classical dynamics for the Hamiltonian~(\ref{Hamilt-HgI2}) of HgI$_2$ versus energy: (a)~The periods of the symmetric-stretch periodic orbit $p=0$ and the asymmetric-stretch periodic orbits $p=1$ and $p=2$ existing above the supercritical pitchfork bifurcation at $E_{\rm c}=523$~cm$^{-1}$. (b)~The Lyapunov exponents $\lambda_p$ of the three periodic orbits, together with the classical escape rate~(\ref{esc}).}
\label{fig4}
\end{figure}

Figure~\ref{fig4} shows the periods and Lyapunov exponents of the three periodic orbits of shortest period, namely, the symmetric-stretch periodic orbit $p=0$ and the two asymmetric-stretch periodic orbits $p=1$ and $p=2$.  We see in Fig.~\ref{fig4}b that their Lyapunov exponents are vanishing at the pitchfork bifurcation.  The Lyapunov exponent of $p=0$ remains equal to zero in the interval $E_{\rm c}<E<548$~cm$^{-1}$ where this orbit is elliptic.  Figure~\ref{fig4}b also shows the classical escape rate~(\ref{esc}), which is equal to the Lyapunov exponent of $p=0$ in the periodic regime for $E<E_{\rm c}$.  In the chaotic regime, the escape rate is smaller than the Lyapunov exponents because of dynamical randomness characterized by the KS entropy in Eq.~(\ref{esc}).  The energy dependences of the period and the coefficient $\chi(E)$ for the symmetric-stretch periodic orbit $p=0$ are well described by the fits
\bea
T(E) &=&  435.57 +  0.17557\, \Delta E +  7.15\times 10^{-5} \, \Delta E^2 \, , \label{T0}\\
\chi(E) &=& 0.13732\, \Delta E + 2.6733 \times 10^{-4} \, \Delta E^2 + 1.9897 \times 10^{-7} \, \Delta E^3 \, , 
\label{chi0}
\eea
with $\Delta E=E-E_{\rm c}$ and the chosen units.  The reduced action of this periodic orbit is obtained by integrating the polynomial fit~(\ref{T0}) over energy:
\be
S(E) = \int_0^E T(E')\, dE' \, .
\ee
A similar fit has been carried out for the common period of the asymmetric-stretch periodic orbits $p=1$ and $p=2$ in order to get with Eq.~(\ref{alpha}) the value of the coefficient
\be
\alpha  = 0.038113/(\mbox{fs cm}^{-1}) \, ,
\label{alpha-value}
\ee
which is positive so that the pitchfork bifurcation is supercritical.

\begin{figure}
\includegraphics[width=0.55\textwidth]{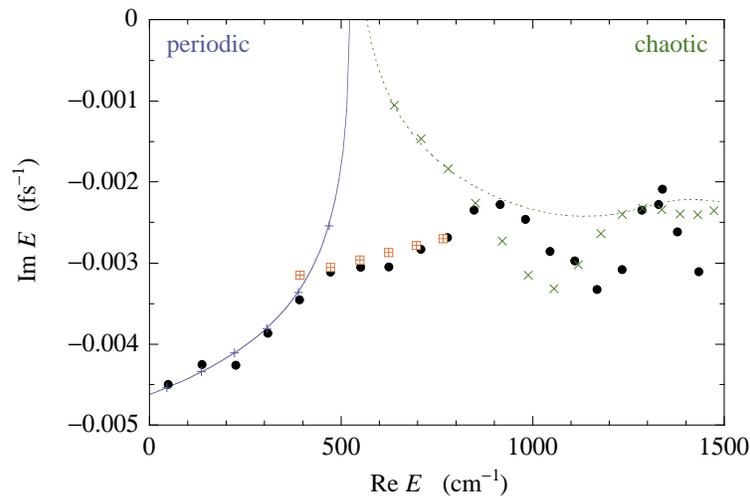}
\caption{Scattering resonances of the Hamiltonian~(\ref{Hamilt-HgI2}) of HgI$_2$ obtained with different methods. The values obtained with the wavefunction method in Ref.~\cite{BG94} are depicted by filled circles. The pluses are the semiclassical approximations given by the zeroes of Eq.~(\ref{quantiz-per}) in the periodic regime.  The solid line corresponds to the gap ${\rm Im}\, E= -\hbar\,\lambda_0({\rm Re}\, E)/2$ in terms of the Lyapunov exponent $\lambda_0$ of the symmetric-stretch periodic orbit $p=0$. The crosses are the semiclassical approximations given by the zeroes of the zeta function~(\ref{Zeta-3}) in the chaotic regime \cite{BG94,BG95}.  In this regime, the dashed line gives the gap ${\rm Im}\, E=-\hbar\,\gamma_{\rm q}({\rm Re}\, E)/2$ in terms of the quantum escape rate~(\ref{esc-q}).  In the intermediate regime, the crossed squares depicts the local approximation here obtained with the formula~(\ref{quantiz-bif-per}).}
\label{fig5}
\end{figure}

The quantum scattering resonances have been calculated by wavepacket propagation using a numerical integration of the Schr\"odinger equation (filled circles in Fig.~\ref{fig5}) \cite{BG94}.  

In the classically periodic regime, the resonances can be obtained semiclassically using Eq.~(\ref{quantiz-per}) (pluses in Fig.~\ref{fig5}).  These resonances are located on the line, ${\rm Im}\, E_n = -\Gamma_n/2 = -\hbar\,\lambda_0({\rm Re}\, E_n)/2$, which is determined by the Lyapunov exponent $\lambda_0(E)$ of the symmetric-stretch periodic orbit $p=0$. This line in Fig.~\ref{fig5} forms the gap~(\ref{esc-q}) in the resonance spectrum.  

In the classically chaotic regime, semiclassical approximations for the resonances are given by the zeroes of the zeta function~(\ref{Zeta-3}) using the characteristic quantities of the three shortest periodic orbits (crosses in Fig.~\ref{fig5}) \cite{BG94}.  In this regime, the gap in the resonance spectrum is evaluated in terms of the quantum escape rate defined by Eq.~(\ref{esc-q}) (dashed line in Fig.~\ref{fig5}) \cite{GB97}.  

We observe that the gap is closed in the intermediate regime between the pitchfork bifurcation at $E_{\rm c}=523$~cm$^{-1}$ and the energy $E\simeq 568$~cm$^{-1}$ where the quantum escape rate~(\ref{esc-q}) becomes positive.  We notice that this energy interval includes the one where the Lyapunov exponents are vanishing.  However, the KAM elliptic island existing in this intermediate regime has an area smaller than Planck's constant $\hbar$, so that the quantum-mechanical wavefunction ignores these nonhyperbolic classical features and escapes nevertheless.

\begin{table}
\caption{Scattering resonances of the Hamiltonian~(\ref{Hamilt-HgI2}) of HgI$_2$.
The values $\{E_n^{\rm (wf)}\}$ are obtained with the wavefunction method in Ref.~\cite{BG94} 
and $\{E_n^{\rm (sc)}\}$ with the formula~(\ref{quantiz-bif-per}).  The imaginary parts of the complex energies are given in terms of the lifetimes by ${\rm Im}\, E_n=-\hbar/(2\tau_n)$ with $\hbar=5308.84$~fs~cm$^{-1}$.}
\label{tab:1}       
\begin{tabular}{lllll}
\hline\noalign{\smallskip}
$n\qquad$ & ${\rm Re}\, E_n^{\rm (wf)}$ (cm$^{-1}$) & $\tau_n^{\rm (wf)}$ (fs) $\qquad\qquad$ & ${\rm Re}\, E_n^{\rm (sc)}$ (cm$^{-1}$) & $\tau_n^{\rm (sc)}$ (fs) \\
\noalign{\smallskip}\hline\noalign{\smallskip}
0   &   47.8 & 111.1 & - & - \\
1   & 136.9 & 117.6 & - & - \\
2   & 225.2 & 117.3 & - & - \\
3   & 308.6 & 129.5 & - & - \\
4   & 391.5 & 144.9 & 393.5 & 158.6 \\
5   & 472.2 & 160.8 & 472.8 & 163.6 \\
6   & 551.2 & 163.8 & 549.8 &  168.7 \\
7   & 624.7 & 164.1 & 624.4 &  174.0 \\
8   & 708.2 & 176.8 & 696.7 &  179.4 \\
9   & 777.9 & 186.2 & 766.9 &  184.9 \\
\noalign{\smallskip}\hline
\end{tabular}
\end{table}

In this intermediate regime, the method of Section~\ref{sec:3} applies.  Using the fits~(\ref{T0})-(\ref{chi0}) and the value~(\ref{alpha-value}) of the coefficient~$\alpha$ in the formula~(\ref{quantiz-bif-per}), a local approximation is obtained for the quantum scattering resonances (crossed squares in Fig.~\ref{fig5} and values in Table~\ref{tab:1}).  The comparison with the values from the wavefunction method shows that the formula~(\ref{quantiz-bif-per}) gives a good approximation for the energy and lifetime of the resonances in the intermediate regime.  Instead of vanishing as expected from the behavior of the Lyapunov exponents, the imaginary part of the complex energies keeps a non-zero value consistent with the results of the numerical integration of Schr\"odinger's equation.

\section{Conclusions}
\label{sec:5}

In this paper, a study is reported of the quantum scattering resonances of dissociating molecules in the intermediate regime where a pitchfork bifurcation happens in the classical dynamics.  

Below this bifurcation, the only classical trajectory existing above the energy barrier is the unstable symmetric-stretch periodic orbit $p=0$.  This orbit becomes stable and surrounded by a small KAM elliptic island at the bifurcation.  This island is bordered by two unstable asymmetric-stretch periodic orbits $p=1$ and $p=2$.  At higher energies, further bifurcations occur leading to the formation of a fully chaotic saddle described by the triadic symbolic dynamics based on the three aforementioned periodic orbits $p=0,1,2$.

In the classically periodic and chaotic regimes, the quantum scattering resonances can be obtained to a good approximation with the standard periodic-orbit theory.  Since this theory makes use of the Lyapunov exponents that are vanishing at the bifurcation, another method is required in the intermediate regime near the bifurcation. Such a method is here developed thanks to a local approximation valid close to the supercritical pitchfork bifurcation. In this way, an analytical formula is obtained for the scattering resonances in the intermediate regime.  This formula is applied to the dissociation of HgI$_2$ where a supercritical pitchfork bifurcation manifests itself.  The results are in agreement with the values of the quantum scattering resonances obtained with the numerical simulation of wavepacket propagation ruled by the Schr\"odinger equation.  

Further work is needed to see if the approximation can be improved and if other bifurcations can be analyzed similarly.

\begin{acknowledgements}
This research is financially supported by the Universit\'e Libre de Bruxelles and the Belgian Federal Government under the Interuniversity Attraction Pole project P7/18 ``DYGEST".
\end{acknowledgements}


\begin{thebibliography}{99}

\bibitem{B81} M. V. Berry, Ann. Phys. {\bf 131}, 163 (1981).

\bibitem{G90} M. C. Gutzwiller, {\it Chaos in Classical and Quantum Mechanics}. Springer-Verlag, New York (1990).

\bibitem{V88} A. Voros, J. Phys. A {\bf 21}, 685 (1988).

\bibitem{GR89a} P. Gaspard and S. A. Rice, J. Chem. Phys. {\bf 90}, 2225 (1989).

\bibitem{GR89b} P. Gaspard and S. A. Rice, J. Chem. Phys. {\bf 90}, 2242 (1989).

\bibitem{GR89c} P. Gaspard and S. A. Rice, J. Chem. Phys. {\bf 90}, 2255 (1989).

\bibitem{CE89} P. Cvitanovi\'c and B. Eckhardt, Phys. Rev. Lett. {\bf 63}, 823 (1989).

\bibitem{ERTW91} G. S. Ezra, K. Richter, G. Tanner, and D. Wintgen, J. Phys. B {\bf 24}, L413 (1991).

\bibitem{GR93} P. Gaspard and S. A. Rice, Phys. Rev. A {\bf 48}, 54 (1993).

\bibitem{GAB95} P. Gaspard, D. Alonso, and I. Burghardt, Adv. Chem. Phys. {\bf 90}, 105 (1995).

\bibitem{GB97} P. Gaspard and I. Burghardt, Adv. Chem. Phys. {\bf 101}, 491 (1997).

\bibitem{BG94} I. Burghardt and P. Gaspard, J. Chem. Phys. {\bf 100}, 6395 (1994).

\bibitem{BG95} I. Burghardt and P. Gaspard, J. Phys. Chem. {\bf 99}, 2732 (1995).

\bibitem{BG97} I. Burghardt and P. Gaspard, Chem. Phys. {\bf 225}, 259 (1997).

\bibitem{AA68} V. I. Arnold and A. Avez, {\it Ergodic Problems in Classical Mechanics}. Benjamin, New York (1968).

\bibitem{DBGZ89} M. Dantus, R. M. Bowman, M. Gruebele, and A. H. Zewail, J. Chem. Phys. {\bf 91}, 7437 (1989).

\bibitem{GRZ90} M. Gruebele, G. Roberts, and A. H. Zewail, Phil. Trans. R. Soc. London Ser. A {\bf 332}, 35(1990).

\bibitem{B92a} E. B. Bogomolny, Chaos {\bf 2}, 5 (1992).

\bibitem{B92b} E. B. Bogomolny, Nonlinearity {\bf 5}, 805 (1992).

\bibitem{W88} S. Wiggins, {\it Global Bifurcations and Chaos}. Springer-Verlag, New York (1988).

\bibitem{W92} S. Wiggins, {\it Chaotic Transport in Dynamical Systems}. Springer-Verlag, New York (1992).

\bibitem{N95} G. Nicolis, {\it Introduction to Nonlinear Science}. Cambridge University Press, Cambridge UK (1995).

\bibitem{GR89JPC} P. Gaspard and S. A. Rice, J. Phys. Chem. {\bf 93}, 6947 (1989).

\bibitem{ER85} J.-P. Eckmann and D. Ruelle, Rev. Mod. Phys. {\bf 57}, 617 (1985).

\bibitem{KG85} H. Kantz and P. Grassberger, Physica D {\bf 17}, 75 (1985).

\bibitem{PP78} P. Pechukas and E. Pollak, J. Chem. Phys. {\bf 69}, 1218 (1978).

\bibitem{CEW11} P. Collins, G. S. Ezra, and S. Wiggins, J. Chem. Phys. {\bf 134}, 244105 (2011).

\bibitem{T00} J. R. Taylor, {\it Scattering Theory: The Quantum Theory of Nonrelativistic Collisions}. Dover, New York (2000).

\bibitem{AB81} A. Bohm, J. Math. Phys. {\bf 22}, 2813 (1981).

\bibitem{R65} L. Rosenfeld, Nucl. Phys. {\bf 70}, 1 (1965).

\bibitem{OH87} A. M. Ozorio de Almeida and J. H. Hannay, J. Phys. A {\bf 20}, 5873 (1987).

\bibitem{AE94} K. M. Atkins and G. S. Ezra, Phys. Rev. A {\bf 50}, 93 (1994).

\bibitem{SS97} H. Schomerus and M. Sieber, J. Phys. A {\bf 30}, 4537 (1997).

\bibitem{AS72} M. Abramowitz and I. A. Stegun, {\it Handbook of Mathematical Functions}. Dover Publications, New York (1972).

\end{thebibliography}
\end{document}